\documentstyle[preprint,aps,12pt]{revtex}
\begin{document}
\title{Ferroelectric thin films properties: depolarization field and
Landau free-energy coefficients renormalization}
\author{M.D.GLINCHUK$^a$, E.A.ELISEEV$^a$, V.A.STEPHANOVICH$^b,$ R.FARHI$^c$}
\address{$^a$Institute for Problems of Materials Science, National Academy\\
of Science of Ukraine, Krjijanovskogo 3, 03142 Kiev, Ukraine \\
$^b$Institute of Mathematics, Opole University, 45-052 Opole, Poland \\
$^c$Laboratory of Condensed Matter Physics, University of Picardy, 800039\\
Amiens, France}
\maketitle

\begin{abstract}
The calculation of the polarization in ferroelectric thin films is performed
using an analytical solution of the Euler-Lagrange differential equation
with boundary conditions with different extrapolation lengths of positive
sign on the surfaces. The depolarization field effect is taken into account
in the model for a short-circuited single domain film, that is a perfect
insulator. It is shown that the calculation of the polarization and other
properties profiles and average values can be reduced to the minimization of
the free energy expressed as a power series of the average polarization with
a renormalized coefficient which depends on temperature, film thickness,
extrapolation lengths, and a coefficient for the polarization gradient term
in the free energy functional, the depolarization field being also included
into the renormalized coefficient. The function defining the space
distribution properties is calculated as well and its amplitude is shown to
coincide with the average polarization. The detailed calculations of the
spontaneous polarization, dielectric susceptibility and pyrocoefficient is
performed. The divergence of the dielectric susceptibility and
pyrocoefficient for critical parameters of the thickness induced
ferroelectric phase transition, namely at temperature $T_{cl}$ and critical
length $l_c$, is shown to exist with and without the depolarization field
contribution, although the values of $T_{cl}$ and $l_c$ are different in
both cases. The detailed analysis of the depolarization field space
distribution and of this field dependence on temperature and film thickness
is performed.
\end{abstract}

\section{Introduction}

The non homogeneity of the polarization and other properties is known to be
a characteristic feature of ferroelectric thin films. The physical reason of
this phenomenon is related to the influence of the films surface, where,
e.g., the polarization differs essentially from that in the bulk. Because of
this polarization heterogeneity, the phenomenological approach to its
calculation was based on the solution of Euler-Lagrange and Lame type
differential equations respectively for polarization and dielectric
susceptibility with specific boundary conditions [1,2]. Because the
analytical solution of these equations appeared to be cumbersome, their
solution was performed numerically in most of the papers, mainly for BaTiO$%
_3 $ and PbTiO$_3$ (see e.g. [3,4,5]). Therefore the analytical calculations
of the film properties appeared to be more complex than in bulk materials,
where the minimization of the free energy as a power series of homogeneous
polarization makes it possible to calculate the properties analytically by a
simple way. The real situation in films is even more complex because of the
depolarization field contribution, which, contrary to the bulk, is non zero
even in short circuited films because of the polarization non homogeneity.
Although the depolarization field is able to completely destroy
ferroelectric polarization [6] there are only a few works devoted to
calculations of its contribution to the film properties [7]. The majority of
the authors performed the calculations without taking into account the
depolarization field effect [3,4,5]. It might be supposed that the
flattening of the polarization via the depolarization field contribution, as
shown in [7], could create conditions for a simplification of the thin film
properties description, particularly on the basis of a free energy expansion
similar to that in bulk materials, but with renormalized coefficients. Such
approach was successfully applied recently to the description of thin films
[8] and nanomaterials [9]. However, up to now, there was no fundamental
background for such a simplified approach. In addition, nothing was
discussed about the properties profiles, which cannot be obtained from
conventional expansion of the free energy as a power series of the
polarization.

In the present work we show for the first time that polarization and other
properties averaged over the film thickness can be found by the minimization
of the free energy with the same form as known for the bulk materials, but
with the coefficient before square polarization depending upon temperature
and film thickness, and featuring the extrapolation length and coefficient
before polarization gradient. The space distribution of the polarization is
calculated using the analytical solution of the Euler-Lagrange equation. The
amplitude of this distribution is shown to be equal to the average
polarization with a good accuracy and can thus be obtained from the
aforementioned free energy. The depolarization field contribution is taken
into account in the model which was proposed in [7] for single domain
ferroelectric films under short circuit conditions, ferroelectric treated as
a perfect insulator. This model seems to be reasonable because the thinner
the film the better the conditions for single domain state appearance (see
e.g. [2], [10]). In addition, for many perovskite ferroelectrics, the
conductivity can be small enough [11]. To analyze the depolarization field
effect we perform a detailed comparison of the properties calculated with
and without taking into account the depolarization field.

\section{Basic equations}

Let us consider a thin ferroelectric film polarized along the $z$-axis (i.e. 
$P_z\neq 0$, $P_x=P_y=0$), which is perpendicular to the surface of the
film. This type of polarization can appear as a result of self-polarization
of a film grown under special technological condition without application of
any external electric field [12,13]. Since the type of substrate and
electrode was shown to be important, the mechanical strain related to the
mismatch of the substrate and the film lattice constants and the thermal
expansion coefficients can contribute to a self-polarization phenomenon.

The polarization equilibrium value can be obtained in the framework of the
phenomenological thermodynamic theory from the minimum of the functional of
the free energy [14]. In the considered case of polarization perpendicular
to the surface of the film, it is necessary to take into account the
depolarization field, which is proportional to the value of polarization and
has the opposite direction, in such a way that it lowers or even cancels $%
P_s $. In accepted models, the ferroelectric is regarded as a perfect
insulator under short-circuit conditions. Taking into account the symmetry
of the considered structure (i.e. the film polarization depends on the
coordinate $z $ only) one can write the free energy density functional for
phase transitions of the 1st and 2nd order ($\gamma \neq 0$ and $\gamma =0$
respectively) as follows:

\begin{eqnarray}
F &=&\frac 1l\int\limits_0^l\left[ \frac \alpha 2P_z^2(z)+\frac \beta 4%
P_z^4(z)+\frac \gamma 6P_z^6(z)+\frac \delta 2\left( \frac{dP_z(z)}{dz}%
\right) ^2-E_zP_z(z)\right] dz+ \\
&&+\frac \delta {2l}\left( \frac{P_z^2(0)}{\lambda _1}+\frac{P_z^2(l)}{%
\lambda _2}\right) +2\pi \left[ \frac 1l\int\limits_0^lP_z^2(z)dz-\left( 
\frac 1l\int\limits_0^lP_z^2(z)dz\right) ^2\right] .  \nonumber
\end{eqnarray}

Here $l$ is the thickness of the film, the coefficient $\alpha $ depends on
temperature $T$ as $\alpha =\alpha _0(T-T_c)$, $T_c$ is the temperature of
the transition from paraelectric to ferroelectric phase in the bulk
ferroelectric, $E_z$ is the external electric field, $\lambda _{1,2}$ are
extrapolation lengths. The last two and the previous two terms represent the
depolarization field energy and surface energy respectively. In what follows
we will consider the case of positive extrapolation lengths, i.e., when
polarization on the surface is smaller then in the center of the film. Only
in this case surface effects can lead to a size-driven phase transition [1].

The minimization of the functional (1) leads to the following equation
determining the space distribution of polarization over the thickness of the
film:

\begin{equation}
\alpha P_z+\beta P_z^3+\gamma P_z^5-\delta \frac{d^2P_z}{dz^2}=E_z-4\pi
P_z+4\pi \frac 1l\int\limits_0^lP_zdz  \eqnum{2a}  \label{2a}
\end{equation}

with boundary conditions on the surfaces of the film:

\begin{equation}
\left. \left( \lambda _1\frac{dP_z}{dz}-P_z\right) \right| _{z=0}=0,\quad
\left. \left( \lambda _2\frac{dP_z}{dz}+P_z\right) \right| _{z=l}=0. 
\eqnum{2b}  \label{2b}
\end{equation}

The last two terms in Eq. (2a) represent the depolarization field:

\begin{equation}
E_z^d=-4\pi \left( P_z-\frac 1l\int\limits_0^lP_zdz\right) .  \eqnum{3}
\label{3}
\end{equation}

Here the first term is the depolarization field for the free standing film,
and the second one is the compensating field of free charges on the
short-circuiting electrodes. The depolarization field (3) is of course zero
when the polarization space distribution is homogeneous.

\section{Critical temperature and thickness of the size-driven phase
transition}

Equation (2a) for the polarization space distribution is a non-linear
inhomogeneous integral differential equation and is cumbersome to solve
analytically. But in the paraelectric phase, where the spontaneous
polarization $P_s$ equals zero and the polarization is only proportional to
the external field, the non-linear terms in Eq. (2a) can be negligibly small.

Therefore the solution of this linear equation can be expressed in terms of
elementary functions. With boundary conditions (2b), the space distribution
of the polarization in this case has the following form:

\begin{eqnarray}
P^{PE}(\xi ) &=&\frac{1-\varphi (\xi )}{4\pi (\Phi -f)}E;  \nonumber \\
\varphi (\xi ) &=&\frac{(1+d_2-(1-d_1)\exp (-h))\exp (-\xi
)+(1+d_1-(1-d_2)\exp (-h))\exp (\xi -h)}{(1+d_1)(1+d_2)-(1-d_1)(1-d_2)\exp
(-2h)};  \eqnum{4}  \label{4} \\
\Phi &=&\left\langle \varphi \right\rangle =\frac{1-\exp (-h)}h\frac{%
2+d_1+d_2-(2-d_1-d_2)\exp (-h)}{(1+d_1)(1+d_2)-(1-d_1)(1-d_2)\exp (-2h)}. 
\nonumber
\end{eqnarray}

Here all '$z$' indexes are omitted, angle bracket means $z$ coordinate
averaging and the following definitions are introduced:

\begin{equation}
\xi =\sqrt{1-f}\frac z{l_0};\quad f=\frac{-\alpha }{4\pi };\quad l_d^2=\frac %
\delta {4\pi };\quad d_i=\sqrt{1-f}\frac{\lambda _i}{l_d};\quad h=\sqrt{1-f}%
\frac l{l_d}.  \eqnum{5}  \label{5}
\end{equation}

The average polarization of the film can be easily derived from (4):

\begin{equation}
\left\langle P^{PE}\right\rangle =\frac{1-\Phi }{4\pi (-f+\Phi )}E. 
\eqnum{6}  \label{6}
\end{equation}

Note that in the case $d_1=d_2$ the space distribution of the polarization
in the paraelectric phase (4) and its average value (6) coincides with those
obtained earlier [7].

The derivative $\chi =\left. (dP/dE)\right| _{E=0}$ represents the linear
dielectric susceptibility of the film. The analysis of expression (6) has
shown that the value of $\chi $ is positive when either $f<0$ (i.e. $\alpha
>0$, $T>T_c$ - paraelectric phase of bulk ferroelectrics) or $f>0$ and $\Phi
>f$ ($T<T_c$- ferroelectric phase of bulk ferroelectrics). The dielectric
susceptibility diverges for $\Phi =f$, and the considered system undergoes a
phase transition.

Because of the $f$ and $\Phi $ dependences on thickness and temperature,
equation $\Phi =f$ determines accordingly the critical temperature $T_{cl}$
at fixed thickness or critical thickness $l_c$ at fixed temperature.

Note that these critical parameters correspond to the points where the
ferroelectric phase loses its stability (1st order phase transition) or to
the critical temperature and thickness for a 2nd order ferroelectric phase
transition. The parameters $f$ and $l_d$ can be estimated respectively as $%
1/\varepsilon $ and $r_c/\sqrt{\varepsilon }$ where $\varepsilon $ and $r_c$
are the dielectric permittivity and correlation radius of the bulk
ferroelectric in the paraelectric phase respectively. Therefore for
conventional values of these quantities $f\ll 1$ and $l/l_d\approx h\gg 1$.
These inequalities essentially simplify the form of $\varphi (\xi )$ and $%
\Phi $ so the equation determining the phase transition point can be
rewritten as follows:

\begin{equation}
h=\frac{U_1+U_2}f,\quad U_i=\frac 1{1+d_i}.  \eqnum{7}  \label{7}
\end{equation}

The critical thickness is readily obtained from this expression:

\begin{equation}
l_c=\frac{U_1+U_2}fl_d=\frac{4\pi }{-\alpha }\sqrt{\frac \delta {4\pi }}%
\left[ \left( \lambda _1\sqrt{\frac{4\pi }\delta }+1\right) ^{-1}+\left(
\lambda _2\sqrt{\frac{4\pi }\delta }+1\right) ^{-1}\right] ,  \eqnum{8}
\label{8}
\end{equation}

and the critical temperature

\begin{equation}
T_{cl}=T_c\left[ 1-\frac{l_0^2(0)}l\left( \frac 1{\lambda _1+l_d}+\frac 1{%
\lambda _1+l_d}\right) \right] ,\quad l_0(0)=\sqrt{\frac \delta {\alpha _0T_c%
}}.  \eqnum{9}  \label{9}
\end{equation}

Here $l_0(0)$ is the correlation length at zero temperature. As it follows
from Eq. (9), if the thickness value is less than $l_c(0)=l_0^2(0)/((\lambda
_1+l_d)^{-1}+(\lambda _2+l_d)^{-1})$, then the critical temperature becomes
negative, i.e. phase transition vanishes.

Therefore the ferroelectric phase transition in a film can be achieved by
changing the film thickness at some fixed temperature or by varying the
temperature of the film with given thickness. As a matter of fact the curve
described by equation (9) determines the phase boundaries between the
paraelectric and ferroelectric phases, i.e. the phase diagram of the film in
temperature - thickness coordinates, which is depicted in Fig.1 by solid
curves for different values of the extrapolation lengths. The critical
thickness values $l_c(0)$ are the intersections of the curves with the
abscissa axis. Namely, at $l<l_c(0)$, the paraelectric phase (polarization $%
P_s=0$) exists in the whole temperature region, while at $l>l_c(0)$ both
ferroelectric phase ($P_s\neq 0$) at $T<T_{cl}$ and paraelectric phase at $%
T>T_{cl}$ are present. It should be noticed that paraelectric and
ferroelectric phases should coexist in the FE region of Fig. 1 for first
order phase transitions. To clear up the effect of the depolarization field
we also conducted calculations without any depolarization field
contribution. The results of the calculations of $T_{cl}$ and $l_c$ for 2nd
order phase transitions are given in the Appendix (Eqs. (A5), (A6)) and in
Fig. 1 (dashed curves).

It is seen from Fig. 1 that the depolarization field decreases the critical
temperature and increases the critical thickness. These effects are stronger
when decreasing the extrapolation lengths values (compare solid and dashed
curves in Fig.1). It is clearly seen, from the inset of Fig. 1, that $%
(T_c-T_{cl})\sim 1/l$ (solid curves), while without the depolarization field
contribution $(T_c-T_{cl})\sim 1/l^2$ (dashed curves). A more detailed
critical thickness dependence on the extrapolation lengths is represented in
Fig. 2. One can see that the greater the $d_1$ and $d_2$ values, the smaller
the $l_c$ value. The latter can be related to the flattening of the
polarization space distribution on increasing the values of the
extrapolation lengths, that results into a critical thickness decrease and
in a critical temperature increase(compare curves 1, 2 and 3 in Fig.1).

When inequalities $\Phi <f$ and $f>0$ are valid, the polarization derived by
equation (4) becomes negative, and the non-linearity in equation (2a) cannot
be neglected.

\section{The free energy}

In the ferroelectric phase (the region $l>l_c(0)$ and $T<T_{cl}$) the non-
linearity should be taken into consideration. The simplest way to take it
into account is the direct variational method. We choose solution (4) as a
trial function and an amplitude factor will be treated as a variational
parameter. The condition of FE existence $h\geq (U_1+U_2)/f$ can be
fulfilled because of small $f$ and $l_d$ values. Allowing for $h\gg 1$ we
will look for a polarization space distribution in the ferroelectric phase
with the following form:

\begin{equation}
P^{FE}(\xi )=P(1-\varphi (\xi )),\quad \varphi (\xi )=\left\{ 
\begin{array}{c}
U_1\exp (-\xi ),\quad 0\leq \xi \ll h/2; \\ 
0,\quad \xi \sim h/2; \\ 
U_2\exp (\xi -h),\quad 0\leq h-\xi \ll h/2.
\end{array}
\right.  \eqnum{10}  \label{10}
\end{equation}

Here $P$ is the variational parameter that represents the amplitude of the
polarization space distribution.

After calculation of integral (1) with the trial function (10), one can
easily obtain the free energy density as follows:

\begin{equation}
F=\alpha \left( 1-\frac{A_1}f\right) (1-A_1)\frac{P^2}2+\beta (1-B)\frac{P^4}%
4+\gamma (1-C)\frac{P^6}6-EP(1-A_1)  \eqnum{11a}  \label{11a}
\end{equation}

Keeping in mind that the polarization average value is $\left\langle
P\right\rangle =P(1-A_1)$one can rewrite Eq.(11a) as:

\begin{equation}
F=\alpha \frac{(1-A_1/f)}{1-A_1}\frac{\left\langle P\right\rangle ^2}2+\beta 
\frac{(1-B)}{(1-A_1)^4}\frac{\left\langle P\right\rangle ^4}4+\gamma \frac{%
1-C}{(1-A_1)^6}\frac{\left\langle P\right\rangle ^6}6-E\left\langle
P\right\rangle .  \eqnum{11b}  \label{11b}
\end{equation}

where the following definitions have been used

\begin{eqnarray}
A_1 &=&\frac{U_1+U_2}h,\quad B=\frac{G(U_1)+G(U_2)}{12h},\quad C=\frac{%
Q(U_1)+Q(U_2)}{60h},  \eqnum{12}  \label{12} \\
G(U_i) &=&48U_i-36U_i^2+16U_i^3-3U_i^4,~  \nonumber \\
Q(U_i) &=&360U_i-450U_i^2+400U_i^3-225U_i^4+72U_i^5-10U_i^6  \nonumber
\end{eqnarray}

It is worth to underline that the terms in Eq. (1) which correspond to
surface energy, depolarizing field and polarization gradient contribute to
the first term in Eqs. (11). Therefore the main peculiarities of thin film
properties have to be related to the coefficient of the second power of the
polarization.

It can be seen that Eqs. (11a) and (11b) have the form of power series of
the amplitude of the polarization space distribution and average
polarization respectively. These equations can be easily rewritten in the
conventional for bulk ferroelectric form

\begin{equation}
F=a\frac{P^2}2+b\frac{P^4}4+c\frac{P^6}6-E^{\prime }P  \eqnum{13a}
\label{13a}
\end{equation}
and

\begin{equation}
F=a_1\frac{\left\langle P\right\rangle ^2}2+b_1\frac{\left\langle
P\right\rangle ^4}4+c_1\frac{\left\langle P\right\rangle ^6}6-E\left\langle
P\right\rangle ,  \eqnum{13b}  \label{13b}
\end{equation}
where the coefficients $a$, $b$, $c$ and $a_1$, $b_1$, $c_1$ can be obtained
from the comparison of Eqs. (11a,b) and (13a,b), namely

\begin{equation}
a=\alpha \left( 1-\frac{A_1}f\right) (1-A_1),\quad b=\beta (1-B),\quad
c=\gamma (1-C),\quad E^{\prime }=E(1-A_1)  \eqnum{14a}  \label{14a}
\end{equation}
and

\begin{equation}
a_1=\frac{\alpha (1-A_1/f)}{1-A_1},\quad b_1=\frac{\beta (1-B)}{(1-A_1)^4}%
,\quad c_1=\frac{\gamma (1-C)}{(1-A_1)^6}.  \eqnum{14b}  \label{14b}
\end{equation}

In the general case, the coefficients (14a,b) depend on the temperature, the
film thickness, the extrapolation lengths and the coefficient before
polarization gradient, as it follows from Eqs. (12), (7) and (5). At first
glance these dependencies seem to be rather complex, although they have been
already simplified due to inequalities $f=1/\varepsilon \ll 1$, $h=l/l_d\gg
1 $. A further simplification of the coefficients (14) was made by rewriting
them via the critical temperature $T_{cl}=T_c-(U_1+U_2)/\alpha _0\cdot 4\pi
l_d/l$ or via the critical thickness $l_c/l_d=(U_1+U_2)4\pi /(\alpha
_0(T_c-T))$ (see the left hand side part of Eq. (8)). The first or the
second type of the coefficients representation can be useful when studying
respectively the temperature (at given thickness) or thickness (at fixed
temperature) dependencies of the film properties. In the first case the
coefficients $a$ in (14a) and $a_1$ in (14b) can be rewritten as:

\begin{equation}
a=\alpha _0(T-T_{cl})\left[ 1-\frac{\alpha _0}{4\pi }(T_c-T_{cl})\right] 
\eqnum{15a}  \label{15a}
\end{equation}
and

\begin{equation}
a_1=\alpha _0(T-T_{cl})\frac 1{1-\frac{\alpha _0}{4\pi }(T_c-T_{cl})}. 
\eqnum{15b}  \label{15b}
\end{equation}

An estimation of the second multipliers in Eqs. (15a) and (15b) has shown
that they are very close to unity (for ferroelectric phase transition of
displacement type $\alpha _0/(4\pi )\leq 10^{-5}$ with $T_c-T_{cl}\leq 10^2$
and for order-disorder type $\alpha _0/(4\pi )\approx 10^{-3}$ with $%
T_c-T_{cl}<10^2$) and so $1-A_1\approx 1$ everywhere in Eqs. (14a) and
(14b). Similar estimations have shown that $B\ll 1$ and $C\ll 1$ so that
with a good accuracy

\begin{equation}
a=a_1=\alpha _0(T-T_{cl}),\quad b=b_1=\beta ,\quad c=c_1=\gamma ,\quad
E^{\prime }=E.  \eqnum{16a}  \label{16a}
\end{equation}

When studying the thickness dependence of the film properties at fixed
temperature, substitution of critical thickness $l_c$(8) into Eq. (14)
yields:

\begin{equation}
a=a_1=K\frac{l_d}{l_c}\left( \frac{l_c}l-1\right) ,\quad K\equiv
(U_1+U_2)4\pi ,  \eqnum{16b}  \label{16b}
\end{equation}
the coefficients $b$, $b_1$, $c$, $c_1$ and $E^{\prime }$ being the same as
in (16a). We should underline once more that the critical temperature $%
T_{cl} $ depends on $l$ and the critical thickness $l_c$ depends on $T$, and
both of them are functions of the extrapolation lengths and parameter $%
\delta $ (see Figs. 1, 2 and Eqs. (8), (9)), the contribution of
depolarizing field also being included. Because of that, all the physical
properties (polarization, dielectric susceptibility, pyrocoefficient,
entropy, thermal capacity etc.), which can be obtained from conventional
minimization of Eqs. (13a), (13b), have to be dependent on the
aforementioned characteristics.

It should be also underlined, that Eqs. (13a), (13b) and (16a), (16b) are
valid both in paraelectric ($T>T_{cl}$, $l<l_c$) and ferroelectric ($%
T<T_{cl} $, $l>l_c$) phases. This is because we chose the trial function for
the description of the polarization profile in the form obtained for the
paraelectric phase (see Eq. (4)) and used the inequality $h\gg 1$, $f\ll 1$,
which are valid practically for any thin film, as discussed earlier.

Since the formulas which express the physical properties via the free energy
coefficients for both first and second order phase transitions are of the
common knowledge (see e.g. [15]), we shall consider only polarization,
dielectric susceptibility and pyrocoefficients with special attention to
their thickness and temperature dependencies, which are the characteristic
features of ferroelectric thin films. For sake of simplicity, we shall
represent, in what follows, these dependencies for the second order phase
transition. Special attention will be payed to depolarization field
contribution to find out the cases and conditions at which these field
effects are more or less important qualititavely or quantitatively.

\section{The specific behavior of the physical properties of ferroelectric
thin films}

\subsection{The space distributions of the physical properties}

The profiles of the physical properties are defined by the function ($%
1-\varphi (\xi )$) in Eq. (10) and their amplitudes can be found by
conventional minimization of free energy (13a) with the coefficients (16a),
(16b). More particularly, for second order phase transitions at $E=0$:

\begin{equation}
P=\sqrt{-\frac{\alpha _0(T-T_{cl})}\beta },~\chi =\frac 1{2\alpha
_0(T_{cl}-T)},~\Pi =\frac{\alpha _0}{\beta 2P}=\frac{\sqrt{\alpha _0}}{\sqrt{%
\beta }2\sqrt{T_{cl}-T}};  \eqnum{17a}  \label{17a}
\end{equation}

\begin{equation}
P=\sqrt{\frac K\beta \frac{l_d}{ll_c}(l-l_c)},~\chi =\frac 1{%
2Kl_d/l_c(1-l_c/l)},~\Pi =\frac{\alpha _0}{2\sqrt{\beta }\sqrt{%
Kl_d/l_c(1-l_c/l)}}  \eqnum{17b}  \label{17b}
\end{equation}
for the ferroelectric phase ($T<T_{cl}$, $l>l_c$) and

\begin{equation}
P=0,~\chi =\frac 1{\alpha _0(T-T_{cl})},~\Pi =0;  \eqnum{18a}  \label{18a}
\end{equation}

\begin{equation}
P=0,~\chi =\frac 1{Kl_d/l_c(l_c/l-1)},~\Pi =0  \eqnum{18b}  \label{18b}
\end{equation}
for the paraelectric phase ($T>T_{cl}$, $l<l_c$).

Here $P$, $\chi $ and $\Pi $ are the amplitudes of the polarization,
dielectric susceptibility and pyrocoefficient which have to be multiplied by
($1-\varphi (\xi )$) to obtain these properties profiles. The formulas
(17a), (18a) and (17b), (18b) can be applied when studying respectively
temperature (at fixed film thickness) and thickness (at fixed temperature)
dependencies.

The space distributions of these quantities are depicted in Figs. 3, 4, 5
for several film thickness at fixed low temperature, $P_{s0}$, $\chi _0$, $%
\Pi _0$ being the quantities value at $l\rightarrow \infty $ (see Eq.
(17b)). One can see that the amplitude of the susceptibility increases much
faster than that of the pyrocoefficient at $l\rightarrow l_c$ while the
amplitude of spontaneous polarization decreases with the decrease of the
film thickness, the shape of all these quantities profiles being completely
the same. It should be underlined that the profiles are asymmetrical for
different extrapolation lengths on the two surfaces. It can be easily seen
that the profiles remain smooth in the most part of the film even for thin
films with thickness close to the critical value $l_c$ (see solid curves 1
and 2). This phenomenon can be explained by the influence of the
depolarization field that tends to flatten the polarization space
distribution as it will be shown later.

\subsection{The average values of the physical properties}

The thickness and temperature dependencies of the properties can be obtained
directly by minimization of the free energy (13b) with respect to Eqs.
(16a), (16b) and accordingly, they can be calculated on the basis of Eqs.
(17) and (18) for the ferroelectric and paraelectric phases respectively
because the average values coincide with the properties profiles amplitude.
As an illustration, we depicted in Figs. 6 and 7 the temperature dependence
of the inverse dielectric susceptibility and pyrocoefficient, the latter
being proportional to the spontaneous polarization (see Eq. (17a)). In Figs.
6, 7 we introduced $P_{s0}(0)=\sqrt{\alpha _0T_c/\beta }$, $\chi
_0(0)=1/(\alpha _0T_c)$ and $\Pi _0=\sqrt{\alpha _0}/(2\sqrt{\beta T_c})$
which are these properties at $T=0$ in the bulk material. The thickness
dependencies are depicted in the insets to Figs. 3, 4, 5. In particular, the
slope of the strait line in the inset to Fig. 4 gives the value of the
parameter $2Kl_d/l_c$ that defined both susceptibility, pyrocoefficients and
polarization (see Eq. (17b) and inset to Fig. 5).

A characteristic feature of the properties temperature dependencies is the
shift of the average properties anomalies to lower temperature on decreasing
the film thickness. Since the divergences of $\chi $ and $\Pi $ are related
to the temperature of the thickness induced ferroelectric phase transition
(see Eqs. (17), (18)), the aforementioned shift is related to the $T_{cl}$
dependence on the film thickness (see Fig. 1). The shift of $T_{cl}$ due to
the depolarization field effect (compare dashed and solid lines in Fig. 1)
defines this field effect on the average properties of the ferroelectric
films. Since the intersections of the curves with abscissa axis in Fig. 1
defines the $l_c(0)/l_0(0)$ value, the shifts of $l_c(0)/l_0(0)$ to larger
values due to the depolarization field contribution is clearly seen.

The estimations of $l_0(0)=\sqrt{\delta /(\alpha _0T_c)}$ and $l_d=\sqrt{%
\delta /(4\pi )}$ values for the parameters of PbTiO$_3$ and BaTiO$_3$ [3,5]
gave respectively $l_0(0)=15$ $\stackrel{\text{o}}{\text{A}}$, $l_d=0,6$ $%
\stackrel{\text{o}}{\text{A}}$ and $l_0(0)=40$ $\stackrel{\text{o}}{\text{A}}
$, $l_d=2$ $\stackrel{\text{o}}{\text{A}}$ which makes it possible to obtain
the $l_c(0)$ value from the intersections in Fig. 1 or by the calculation of 
$l_c$ on the basis of Eq. (8). For instance, for extrapolation lengths $%
\lambda _1=\lambda _2=l_0(0)$ $l_c(0)=30$ $\stackrel{\text{o}}{\text{A}}$
and 80 $\stackrel{\text{o}}{\text{A}}$ with depolarization field
contribution and without it $l_c(0)=24$ $\stackrel{\text{o}}{\text{A}}$ and
63 $\stackrel{\text{o}}{\text{A}}$ for PbTiO$_3$ and BaTiO$_3$ respectively.
Critical thickness has to increase with decrease of $\lambda _1$ and $%
\lambda _2$ values and at $T\neq 0$ namely $l_c(T)=l_c(0)/(1-T/T_c)$ (see
Eq. (8)). Because of the scattered $\delta $ values given in different
works, the obtained $l_c$ values should be considered as estimations only.
But it is obvious that the larger $T_c$ the smaller $l_c$.

An estimation of the $T_{cl}$ value from Fig. 1, e.g. for $l/l_0(0)=10$ and
parameters of curve 2, gives $T_{cl}=0,8T_c$ for the films with thickness
200-500 $\stackrel{\text{o}}{\text{A}}$. Therefore for such thin films the
shift $\Delta T_c=T_c-T_{cl}=0,2T_c$ and, e.g., for BaTiO$_3$ and PbTiO$_3$, 
$\Delta T_c=24^o$ C and $\Delta T_c=108^o$ C respectively, which means that $%
T_{cl}$ can be rather close to the $T_c$ value. For thicker films $T_{cl}$
is expected to be even closer to $T_c$ (see Fig. 1).

\subsection{The depolarization field and its influence on the physical
properties}

The depolarization field is given by Eq. (3), which after substitutions $%
P_z=P(1-\varphi (\xi ))$ and $\left\langle P\right\rangle =P(1-A_1)$ (see
section 4) can be rewritten as 
\begin{equation}
E_z^d(\xi )=-4\pi P(A_1-\varphi (\xi )).  \eqnum{19}  \label{19}
\end{equation}

Although $A_1$ value is small, it cannot be neglected in the region where $%
\varphi (\xi )=0$ (see Eq. (10)). In particular in the very center of the
film: 
\begin{equation}
E_z^d\equiv E_z^d(\xi =h/2)=-4\pi PA_1.  \eqnum{20a}  \label{20a}
\end{equation}

Allowing for $A_1=l_d(U_1+U_2)/l$ one can rewrite Eq. (20a) with respect to
(17b) as: 
\begin{equation}
E_z^d=-K\frac{l_d}l\sqrt{\frac K\beta \frac{l_d}{l_c}\left( \frac{l_c}l%
-1\right) }  \eqnum{20b}  \label{20b}
\end{equation}
or 
\begin{equation}
E_z^d=-\sqrt{\frac{\alpha _0}\beta (T_{cl}-T)}\alpha _0(T_c-T_{cl}). 
\eqnum{20c}  \label{20c}
\end{equation}

Equations (20b) and (20c) should be applied when considering $E_z^d$
thickness or temperature dependencies respectively.

It is clearly seen that $E_z^d=0$ at $T=T_{cl}$ for arbitrary $l$ and $l=l_c$
and $l\rightarrow \infty $ for arbitrary $T$. Between $l=l_c$ and $%
l\rightarrow \infty $, the absolute value of $E_z^d$ achieves a maximum at $%
l_m=\frac 32l_c$. The value of depolarization field at this point is equal
to $E_m^d=\frac 2{3\sqrt{3}}\alpha \sqrt{-\alpha /\beta }=E_{c0}$, where $%
E_{c0}$ is the coercive field of the bulk material. Near the surfaces, when $%
\varphi (\xi )\neq 0$ (see Eq. (10)), the small quantity $A_1$ can be
neglected and $E_z^d$ becomes thus positive, its profile being defined by
the $\varphi (\xi )$ dependence. The point $\xi _{cr}$ at which the change
of depolarization field sign takes place can be obtained easily from Eq.
(19), with $E_z^d(\xi _{cr})=0$ at $\varphi (\xi _{cr})=A_1$. With respect
to Eq. (10) this gives $(1+U_2/U_1)l_d/l=\exp (-\xi _{cr})$ or $-\xi
_{cr}=\ln (l_d/l)+\ln (1+U_2/U_1)$. Keeping in mind that $l/l_d=h\gg 1$, and
that $U_1$ and $U_2$ are quantities of the same order of magnitude (e.g. $%
U_2/U_1=1$ at $\lambda _1=\lambda _2$) one can conclude that for symmetrical
boundary conditions 
\begin{equation}
\xi _{cr}=\ln \frac l{2l_d}.  \eqnum{21}  \label{21}
\end{equation}
Therefore at $\xi <\xi _{cr}$ $E_z^d(\xi )>0$, while $E_z^d(\xi )<0$ at $\xi
>\xi _{cr}$. Similar conditions on the right hand side of the film can be
obtained by substitution of $(h-\xi _{cr})$ for $\xi _{cr}$ in Eq. (21).

To illustrate the $E_z^d$ behaviour we have represented its profile and the
thickness dependence of $E_z^d(z=l/2)$ in the inset b) of Fig. 8. One can
see that, because there is no depolarization field in the film center at
critical thickness, $E_z^d$ increases on increasing $l/l_c$, achieves
maximum at $l_c/l_m=\frac 23$ and at $l\rightarrow \infty $ $%
E_z^d\rightarrow 0$ again. More details for the region $E_z^d(\xi )<0$ are
given in inset a).

The effect of the depolarization field is clearly shown in Figs. 3 to 5. The
dashed lines are the profiles calculated without any depolarization field
contribution following the way described in Appendix for the thinnest (curve
1) and the thickest (curve 5) films. It can be seen that this field
influence depends on the film thickness: the thinner the film the stronger
the change in the profile shape. Comparison of solid and dashed curves in
Figs. 3, 4, 5 shows that the depolarization field increases $\chi $ and $\Pi 
$ values, but decreases the $P$ value for the thin film close to the
critical thickness, while for thick films, a decrease in $\chi $ and $\Pi $
and an increase in $P_s$ are observed in most part of the film (see curve
5). It is worth to notice that the dashed curves in Figs. 4, 5 display small
maxima near the surfaces contrary to solid curves which remains flat. It is
clear that the influence of the depolarization field on the polarization is
directly related to the peculiarities of depolarisation field profile (see
Fig. 8). In particular the positive sign of $E_z^d(\xi )$ in the tiny
regions close to the surfaces and negative sign outside these regions
explains the increase of the spontaneous polarization near the surfaces and
its decrease in most part of the film, leading to the flattening of the
polarization (see Fig. 3). The difference in the behaviors induced by the
depolarization field effect on $P_s$ on one hand, and $\chi $ and $\Pi $,
which are the derivative of $P_s$, on the other hand, is not surprising,
since, e.g., $\Pi \sim 1/P_s$, and an increase in $P_s$ should result in a $%
\Pi $ decrease and vice versa.

\section{Discussion and conclusion}

Let us discuss briefly the accuracy of the calculations and the physical
model we used here. The variational method used for the calculation of the
polarization in the ferroelectric phase is an approximation, its accuracy
being usually a few percent. To prove this we performed the calculations in
two important cases by other exact ways: in the vicinity of the phase
transition, where $P_s$ tends to zero and can be considered as a small
parameter, and for the thick film limit, when the polarization space
distribution is almost uniform. The solutions obtained by these methods are
reported in the inset of Figs. 3, 4, 5 by dashed and dotted-dashed lines. It
can be seen that these lines coincide with those obtained by the variational
method with a good accuracy.

In our calculations we did not take into account the possible shift of $T_c$
by mechanical strains originating from the misfit between the substrate and
the film lattice constants, thermal expansion coefficients and growth
imperfections. Because of the strain relaxation due to misfit dislocations,
the strains are not homogeneous and in the general case their value are
small in most part of the film [16]. We accordingly neglected the
renormalization of $T_c$ value by mechanical strains.

Let us discuss briefly the depolarization field effect. As it follows from
the above considerations, the depolarization field does not change the
general form of $P_s$, $\chi $ and $\Pi $ average values dependences on the
film thickness and temperature (see Figs. 6, 7). The depolarization field
smoothes the space distributions of these quantities, and in particular
cancels the maxima near the film surfaces in the space distributions of $%
\chi $ and $\Pi $ obtained in the case where the depolarization field can be
neglected (see Figs. 4, 5).

The main effect of the depolarization field is the shift of the boundary
between the paraelectric and ferroelectric phases on the temperature - film
thickness phase diagram to larger thickness, as can be seen from Fig. 1.
This shift can be especially large for thin films because the critical
temperature $T_{cl}$ obeys a $T_c-T_{cl}\sim 1/l$ dependence when taking
into account depolarization field contribution (see Eq. (9) and solid lines
in Fig. 1) but a $T_c-T_{cl}\sim 1/l^2$ without this contribution (see Eqs.
(A6a), (A6b) and dashed lines in Fig. 1). Because of that a scattering of
experimental critical values can be expected for films produced in different
technological conditions and from materials with different degrees of
purity. These factors might influence the domain structure and conductivity
of the film and consequently the depolarization field.

Obviously depolarization field effects in thick films under short-circuit
conditions can be neglected in most part of the film except close to the
surfaces. But another important case where the contribution of the
depolarization field can be neglected is related to the conditions on the
film surfaces. More particularly, for extrapolation lengths much larger than
the correlation length $l_0$ it follows from Eq. (A6c) that $T_c-T_{cl}\sim
1/l$ and Eq. (A6c) looks like Eq. (9).

Although we considered spontaneous polarization and other properties without
Any external electric field application, the case of $E\neq 0$ can be
considered on the basis of Eq. (13), i.e., in the same way as nonlinear
effects in the bulk. It should be mentioned that a non linear dielectric
response was measured in Ba$_{0,7}$Sr$_{0,3}$TiO$_3$ thin films with
thickness 24-160 nm [8]. Authors of this work explained the observed
temperature and $E$-field dependencies using a conventional power series
expansion of the free energy. They came to the conclusion that only the
first term of this power series varies significantly with the film thickness
or the temperature. This conclusion is in very good agreement with the
results obtained in our work. The spatial distribution of the pyroelectric
coefficient measured by LIMM method in homogeneously poled ferroelectric
polymer films [17] looks like that depicted in Fig. 5. This may give some
support to the relevance of the calculated profiles of polarization and
dielectric susceptibility.

The obtained results pave the way to calculations of ferroelectric film
properties using a conventional minimization of the free energy for bulk
material, but with renormalized coefficient before squared polarization.
This renormalized coefficient includes the contribution of polarization
gradient, depolarization field, polarization at the surface and the
characteristics of the bulk material. Because this coefficient has been
shown to depend on the critical parameters of the thickness induced
ferroelectric phase transition $T_{cl}$ and $l_c$ in the same way as this
coefficient temperature dependence in the bulk, namely, for the films, $%
a=\alpha _0(T-T_{cl})$ or $a=Kl_d/l_c(l_c/l-1)$ (see Eq. (16)), this makes
it possible to obtain these critical parameters through the observation of
thermodynamic properties anomalies, e.g. by measurements of the average
dielectric susceptibility maximum position at different temperatures at
fixed thickness or at fixed temperature and different film thickness
respectively. The novel approach developed here is suitable for the
calculation of the profiles and average values of the polarization,
dielectric susceptibility, pyrocoefficient, entropy, specific heat etc. In
order to calculate the electromechanical properties, we have to include into
the free energy functional the interaction of the non homogeneous
polarization with external strains. This consideration is in progress now.

\section{Appendix}

In order to discuss the influence of the depolarization field, let us
consider briefly the model without taking into account that field. In this
case the equation giving the space distribution of polarization as a
function of the thickness of the film has the form Eq. (2a) without the last
two terms. This equation has a nontrivial solution when the electric field
is zero. In the case of boundary conditions (2b) with positive extrapolation
lengths, this solution exists only under the condition $\alpha <0$ (see,
e.g. \cite{G-R}) and gives the following space distribution of the
spontaneous polarization for the phase transition of the second order: 
\begin{equation}
P_s(z)=P_{s0}\sqrt{\frac{2m}{1+m}}sn\left( \left. \frac{z+z_0}{l_0\sqrt{1+m}}%
\right| m\right)   \eqnum{A1}  \label{A1}
\end{equation}
Here $l_0$= $\sqrt{-\lambda /\alpha }$ is the correlation length, $sn(u|m)$
is the elliptic sine function \cite{G-R}, and the constants $m$ and $z_0$
have to be determined from the boundary conditions (2b). Using the
properties of the elliptic functions it is easy to obtain equations which
these constants satisfy: 
\begin{equation}
z_0=l_0\sqrt{1+m}~F(\arcsin (f_1),m)  \eqnum{A2}  \label{A2}
\end{equation}
\begin{equation}
l=l_0\sqrt{1+m}\left( 2K(m)-F(\arcsin (f_1),m)-F(\arcsin (f_2),m)\right)  
\eqnum{A3}  \label{A3}
\end{equation}
where $K(m)$ and $F(\phi ,m)$ are complete and incomplete elliptic integrals
of the first kind respectively \cite{G-R}, and the following definition is
introduced: 
\begin{equation}
f_i\equiv f(m,\lambda _i)=\sqrt{\frac{1+m}{2m}\left( 1+\left( \frac{l_0}{%
\lambda _i}\right) ^2-\sqrt{\left( 1+\left( \frac{l_0}{\lambda _i}\right)
^2\right) ^2-\frac{4m}{\left( 1+m\right) ^2}}\right) }  \eqnum{A4}
\label{A4}
\end{equation}
The parameter m varies from 0 to 1 in Eqs. (A1) - (A4), otherwise the
spontaneous polarization would be a complex number. As it follows from Eq.
(A3) with the condition $m\rightarrow 0$, thickness $l$tends to some
critical value $l_c$ which corresponds to a zero value of the spontaneous
polarization (A1). When the thickness of the film is less than critical, the
spontaneous polarization vanishes. The critical thickness for the considered
case $\lambda _1\geq 0$, $\lambda _2\geq 0$ has the form: 
\begin{equation}
l_c=l_0\left( \pi -arctg\left( \frac{\lambda _1}{l_0}\right) -arctg\left( 
\frac{\lambda _2}{l_0}\right) \right)   \eqnum{A5}  \label{A5}
\end{equation}
The critical temperature $T_{cl}$ for a film with thickness $l$ can be found
from the condition $l=l_c(T=T_{cl})$, which results in a transcendental
equation for the dependence of $T_{cl}$ on the film parameters. This
equation can be simplified in several limiting cases. When the phase
transition temperature is much smaller than in the bulk ferroelectric ($%
T_{cl}\ll T_c$), i.e., for very thin films, the following dependence can be
obtained: 
\begin{equation}
T_{cl}=T_c\left( 1-\left( \frac{l_c(0)+l_0(0)\left( g_1+g_2\right) }{%
l+l_0(0)\left( g_1+g_2\right) }\right) ^2\right) ,\quad l-l_c(0)\ll l_0(0) 
\eqnum{A6a}  \label{A6a}
\end{equation}
Here $l_c(0)$ and $l_0(0)=\sqrt{\lambda /(\alpha _0T_c)}$ are the critical
thickness and correlation length respectively at zero temperature $T=0$, and 
$g_i=\lambda _i/\sqrt{l_0^2(0)+\lambda _i^2}.$

When the phase transition temperature is of the same order of magnitude as
in the bulk ferroelectric ($T_{cl}\approx T_c$), i.e. for films with large
thickness, it is easy to get from Eq.(A5) the following expression: 
\begin{equation}
T_{cl}=T_c\left( 1-\left( \frac{\pi l_0(0)}{l+\lambda _1+\lambda _2}\right)
^2\right) ,\quad l\gg l_0(0)  \eqnum{A6b}  \label{A6b}
\end{equation}

Note that this expression is also valid in the whole temperature and
thickness ranges for extrapolation lengths much smaller then the correlation
length ($\lambda _i\ll l_0$). In the opposite case, i.e. $\lambda _i\gg l_0$%
, the critical temperature is expressed as: 
\begin{equation}
T_{cl}=T_c\left( 1-\frac{l_0^2(0)}l\left( \frac 1{\lambda _1}+\frac 1{%
\lambda _2}\right) \right) ,\quad \lambda _i\gg l_0  \eqnum{A6c}  \label{A6c}
\end{equation}

Since $\alpha =\alpha _0(T-T_c)$, where $T_c$ is the temperature of the
ferroelectric phase transition of the thick film, the nonhomogeneous
polarization of the film $P(z)$ should depend on temperature $T$ and
external electric field $E$. This allows to calculate the non homogeneous
pyrocoefficient $\Pi (z)=(dP(z)/dT)_{E=0}$ and the linear dielectric
susceptibility $\chi (z)=(dP(z)/dE)_{E=0}$. Differentiation of Eqs. (2a) and
(2b) gives the following differential equations for the pyrocoefficient and
dielectric susceptibility calculations: 
\begin{equation}
(\alpha +3\beta P_s^2)\Pi -\delta \frac{d^2\Pi }{dz^2}+\alpha _0P_s=0; 
\eqnum{A7a}  \label{A7a}
\end{equation}
\begin{equation}
\left. \frac{d\Pi }{dz}\right| _{z=0}=\left. \frac \Pi {\lambda _1}\right|
_{z=0};\left. \frac{d\Pi }{dz}\right| _{z=l}=-\left. \frac \Pi {\lambda _2}%
\right| _{z=l}  \eqnum{A7b}  \label{A7b}
\end{equation}
\begin{equation}
(\alpha +3\beta P_s^2)\chi -\delta \frac{d^2\chi }{dz^2}-1=0;  \eqnum{A8a}
\label{A8a}
\end{equation}
\begin{equation}
\left. \frac{d\chi }{dz}\right| _{z=0}=\left. \frac \chi {\lambda _1}\right|
_{z=0};\left. \frac{d\chi }{dz}\right| _{z=l}=-\left. \frac \chi {\lambda _2}%
\right| _{z=l}  \eqnum{A8b}  \label{A8b}
\end{equation}
The general solution of Eq. (A8a), which is a particular case of the Lame
equation, is the following \cite{Erd}: 
\begin{equation}
\Pi (x)=C_1^{py}y_1(x)+C_2^{py}y_2(x)+y_3^{py}(x)  \eqnum{A9}  \label{A9}
\end{equation}
where the function $y_i(x)$ has the following form: 
\begin{equation}
y_1(x)=cn(x,m)dn(x,m),  \eqnum{A10a}  \label{A10a}
\end{equation}
Here $cn(u|m)$ and $dn(u|m)$ are the elliptic cosine and delta amplitude
functions respectively \cite{G-R} 
\begin{equation}
y_2(x)=\left( x-\frac{1+m}{1-m}E(am(x),m)\right) \frac{y_1(x)}{1-m}+y_0(x)%
\frac{1+m^2-m(1+m)\,y_0^2(x)}{\left( 1-m\right) ^2},  \eqnum{A10b}
\label{A10b}
\end{equation}
\begin{eqnarray}
y_3^{py}(x) &=&-\Pi _m\left( \left( x-\frac{2\,E(am(x),m)}{1-m}\right)
y_1(x)+y_0(x)\frac{1+m-2m\,\,y_0^2(x)}{1-m}\right)  \nonumber  \label{18} \\
\Pi _m &=&\Pi _0\frac{\sqrt{2m\left( 1+m\right) }}{1-m},\quad \Pi _0=\frac{%
dP_0}{dT}.  \eqnum{A10c}  \label{A10c}
\end{eqnarray}
Here $\Pi _0$ is the thick film pyroelectric coefficient and the following
definitions are introduced: 
\[
x=\frac{z+z_0}{l_0\sqrt{1+m}},\quad y_0(x)=sn(x,m). 
\]
In Eqs.(A10b, c) $E(\varphi ,m)$ and $am(x)$ are an incomplete elliptic
integral of the second kind and an elliptic amplitude function respectively 
\cite{G-R}. The constants $C_1^{py}$ and $C_2^{py}$ are omitted because of
their cumbersome expression.

The dielectric susceptibility can be obtained in a similar way: 
\begin{equation}
\chi (x)=C_1^{ch}y_1(x)+C_2^{ch}y_2(x)+y_3^{ch}(x)  \eqnum{A11}  \label{A11}
\end{equation}
where 
\begin{equation}
y_3^{ch}(x)=-2\chi _0(1+m)\frac{1+m-2m\,y_0(x)}{(1-m)^2},  \eqnum{A12}
\label{A12}
\end{equation}
Here $\chi _0=-1/2\alpha $ is the thick film susceptibility. Coefficients $%
C_1^{ch}$ and $C_2^{ch}$ differ from $C_1^{py}$ and $C_2^{py}$ in Eq.(14)
because of the difference between $y_3^{py}(x)$ and $y_3^{ch}(x)$.

\begin{figure}
\caption{ The dependence of the critical temperature $T_{cl}$ on the film
thickness (logarithmic scale) for $f(0)=\alpha _0T_c/4\pi =0.01$ and for
different extrapolation lengths values: $(\lambda _1/l_d,\lambda
_2/l_d)=(0.5,0.6)$, (5, 3), (10, 6) for lines 1, 2 and 3 respectively with
(solid lines) and without (dashed lines) the depolarization field
contribution. The behavior as a function of the reciprocal film thickness is
given in the inset.}
\end{figure}

\begin{figure}
\caption{The dimensionless critical thickness $fl_c/l_d$ in the $(d_1,d_2)$
space. The $d_i$'s are the dimensionless extrapolation lengths. The critical
thickness is given as numbers close to the relevant plots.}
\end{figure}

\begin{figure}
\caption{The space distribution of the spontaneous polarization $P_s$ as a
function of the $z$ coordinate for the following values of parameters: $%
f=0.01$, $\lambda _1/l_d=5$, $\lambda _2/l_d=3$, and for different values of
the film thickness $l/l_c=1.01$, 1.04, 1.16, 1.7, 4 (solid curves 1 - 5
respectively). The dashed curves 1 and 5 represent the space distribution of 
$P_s$ calculated without taking into account the depolarization field. The
average value of $P_s^2$ vs the reciprocal film thickness $l$ is represented
in the inset. The solid line represents the dependence obtained using the
variational method, dashed and dash-dotted curves are obtained by expansions
close to the phase transition and for the thick films respectively.}
\end{figure}

\begin{figure}
\caption{The space distribution of the dielectric susceptibility in the
ferroelectric phase as a function of the $z$ coordinate for the same values
of parameters as those in Fig. 3. The dashed curves 1 and 5 represent the
dielectric susceptibility space distribution calculated without taking into
account the depolarization field. The inset gives the dependence of the
reciprocal averaged dielectric susceptibility as a function of the
reciprocal thickness of the film. The meaning of the solid, dashed and
dash-dotted curves is the same as in Fig.3.}
\end{figure}

\begin{figure}
\caption{The space distribution of $\Pi $ as a function of the $z$
coordinate for the same values of parameters as those in Fig. 3. The dashed
curves 1 and 5 represent the $\Pi $ space distribution calculated without
taking into account the depolarization field. The inset gives the dependence
of the reciprocal averaged $\Pi $ as a function of the reciprocal thickness
of the film. The meaning of the solid, dashed and dash-dotted curves is the
same as in Fig.3.}
\end{figure}

\begin{figure}
\caption{The temperature dependence of the reciprocal averaged dielectric
susceptibility for the following values of parameters: $f(0)=\alpha
_0T_c/4\pi =0.01$, $\lambda _1/l_d=5$, $\lambda _2/l_d=3$, and for different
values of the film thickness $l/l_c(0)=1.11$, 1.43, 2, 3.33, 10 (curves 1 -
5 respectively). The dashed line depicts the temperature dependence of the
bulk ferroelectric reciprocal dielectric susceptibility $\chi _0(T)$. The
temperature dependence of the average dielectric susceptibility for the same
values of parameters is given in the inset, where the dashed line represents
the bulk ferroelectric dielectric susceptibility.}
\end{figure}

\begin{figure}
\caption{The temperature dependence of the averaged spontaneous
polarization for the same values of parameters as those in Fig. 6. The
temperature dependence of the bulk ferroelectric spontaneous polarization $%
T/T_c$ is plotted as a dashed line. The temperature dependence of the
averaged $\Pi $ for the same values of parameters is given in the inset,
where the dashed line represents the bulk ferroelectric pyroelectric
coefficient $\Pi _0=dP_{s0}/dT=\Pi _0(0)\sqrt{1-T/T_c}$.}
\end{figure}

\begin{figure}
\caption{The space distribution of the depolarization field close to the
film surface for the following values of parameters: $f=0.01$, $\lambda
_1/l_d=5$, $\lambda _2/l_d=3$, and for different values of the film
thickness $l/l_c=1.01$, 1.1, 1.5, 4 (curves 1 - 4 respectively). The region
of the plot where the depolarization field is negative is depicted in the
inset (a). The depolarization field in the center of the film as a function
of the reciprocal film thickness is depicted in the inset (b) for the
parameters values given above.}
\end{figure}

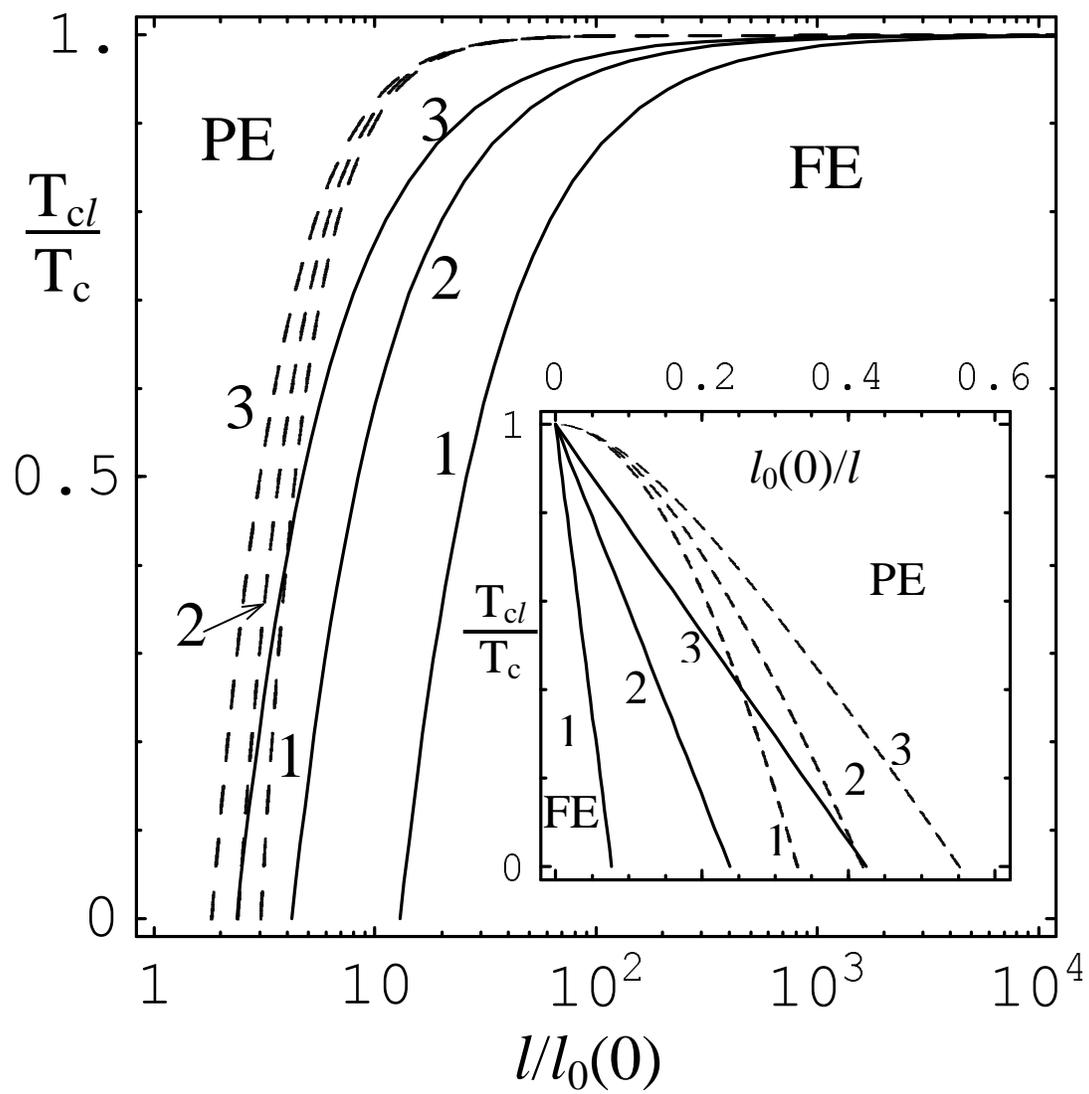
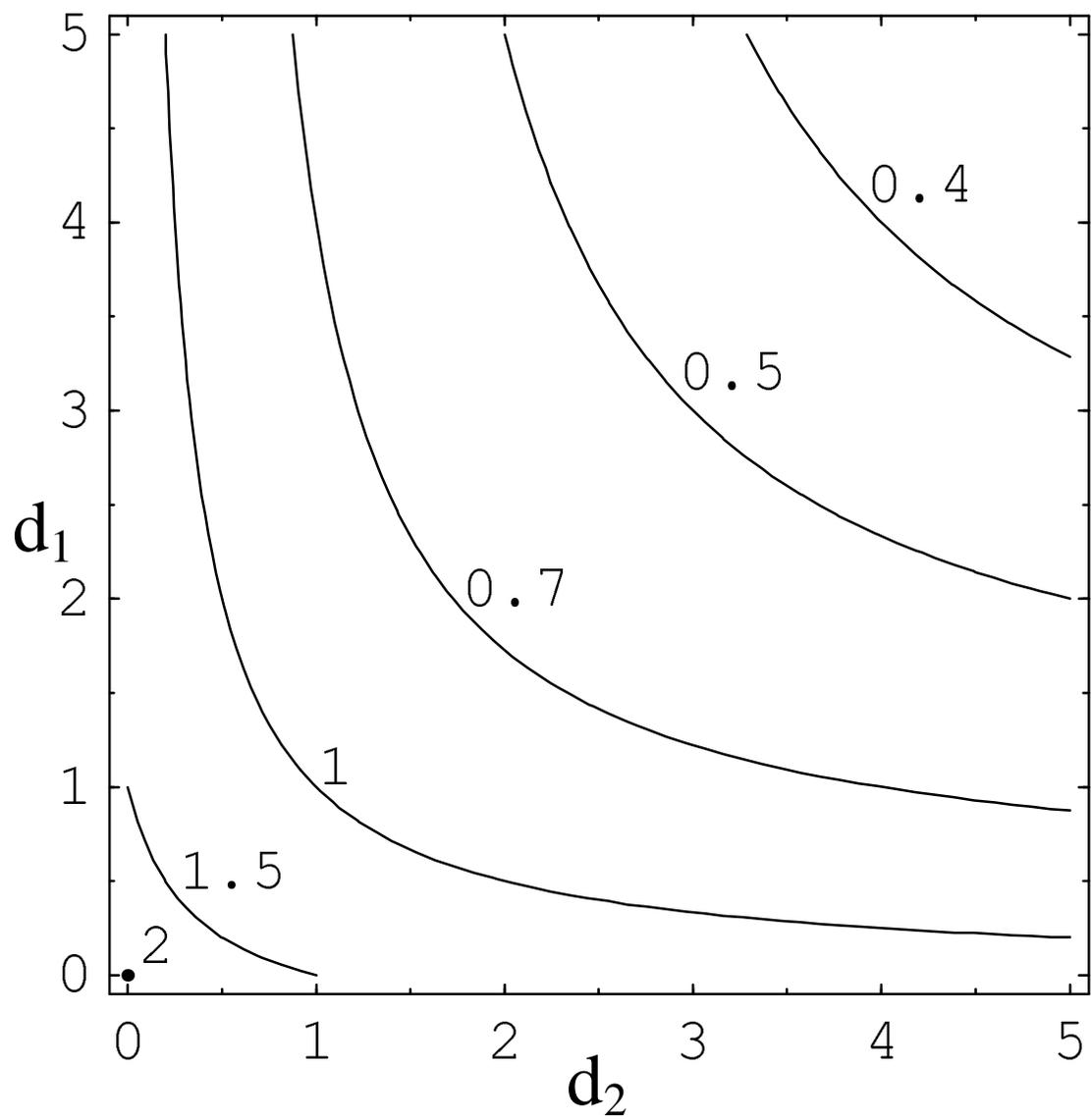
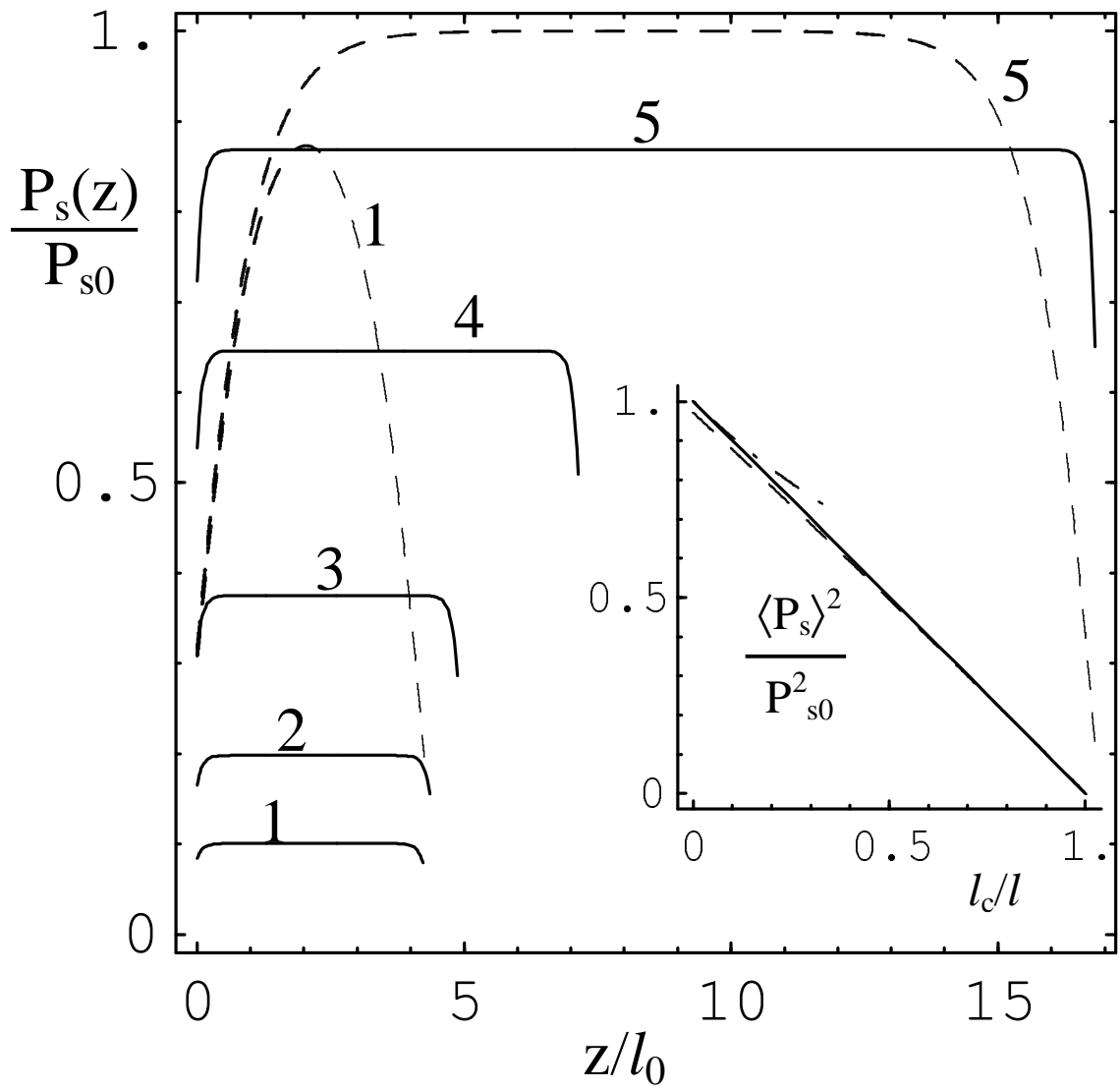
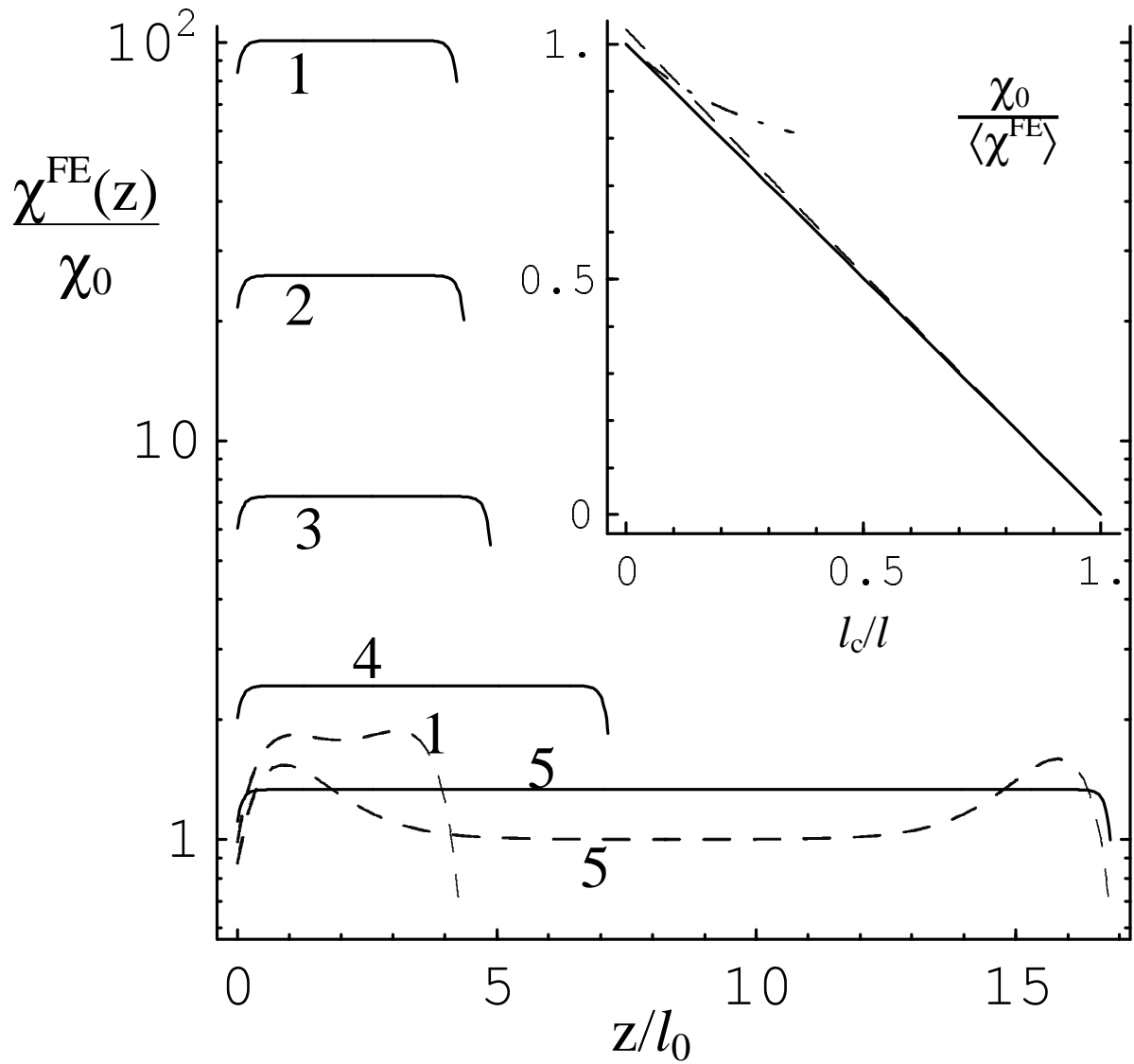
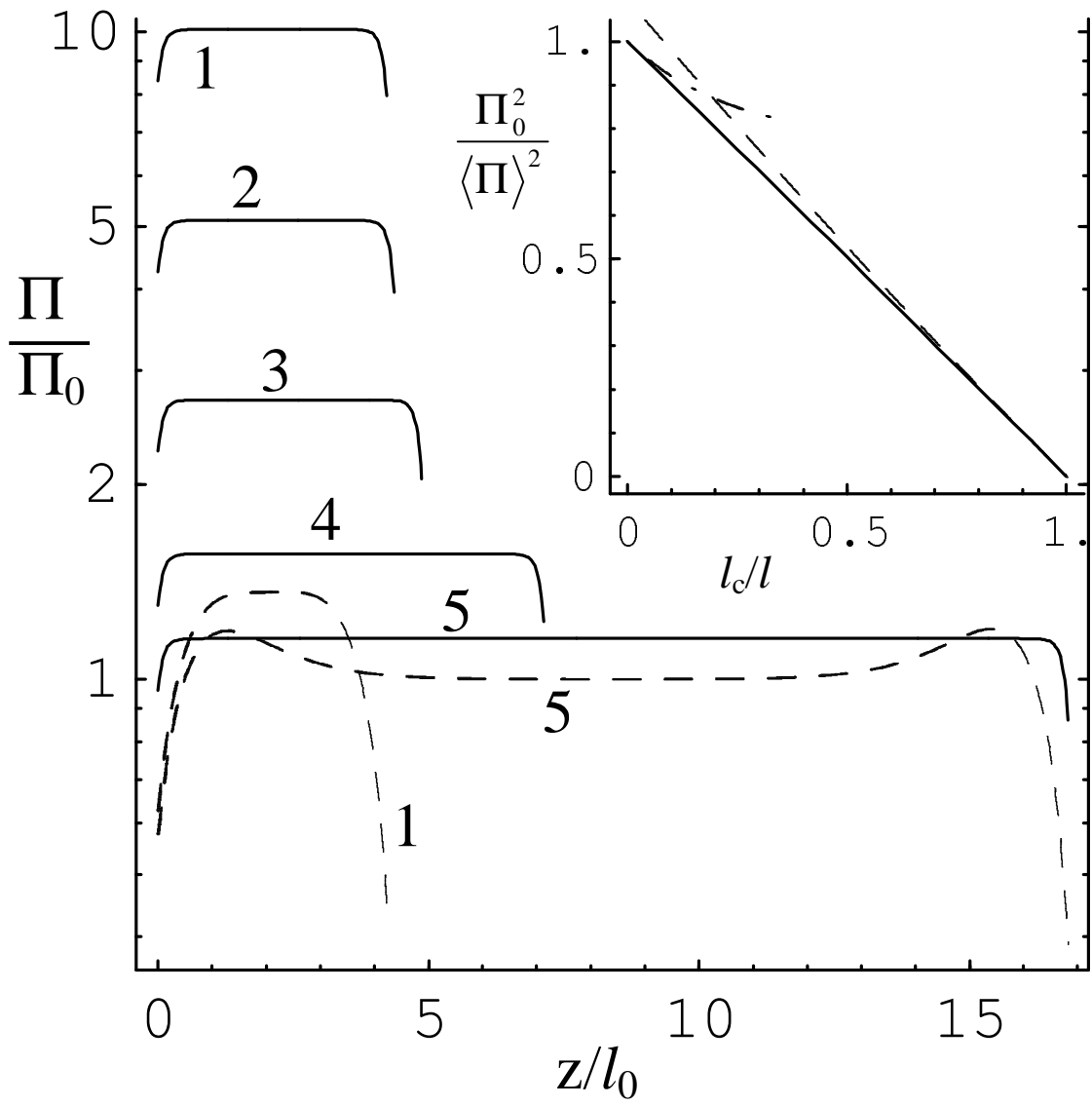
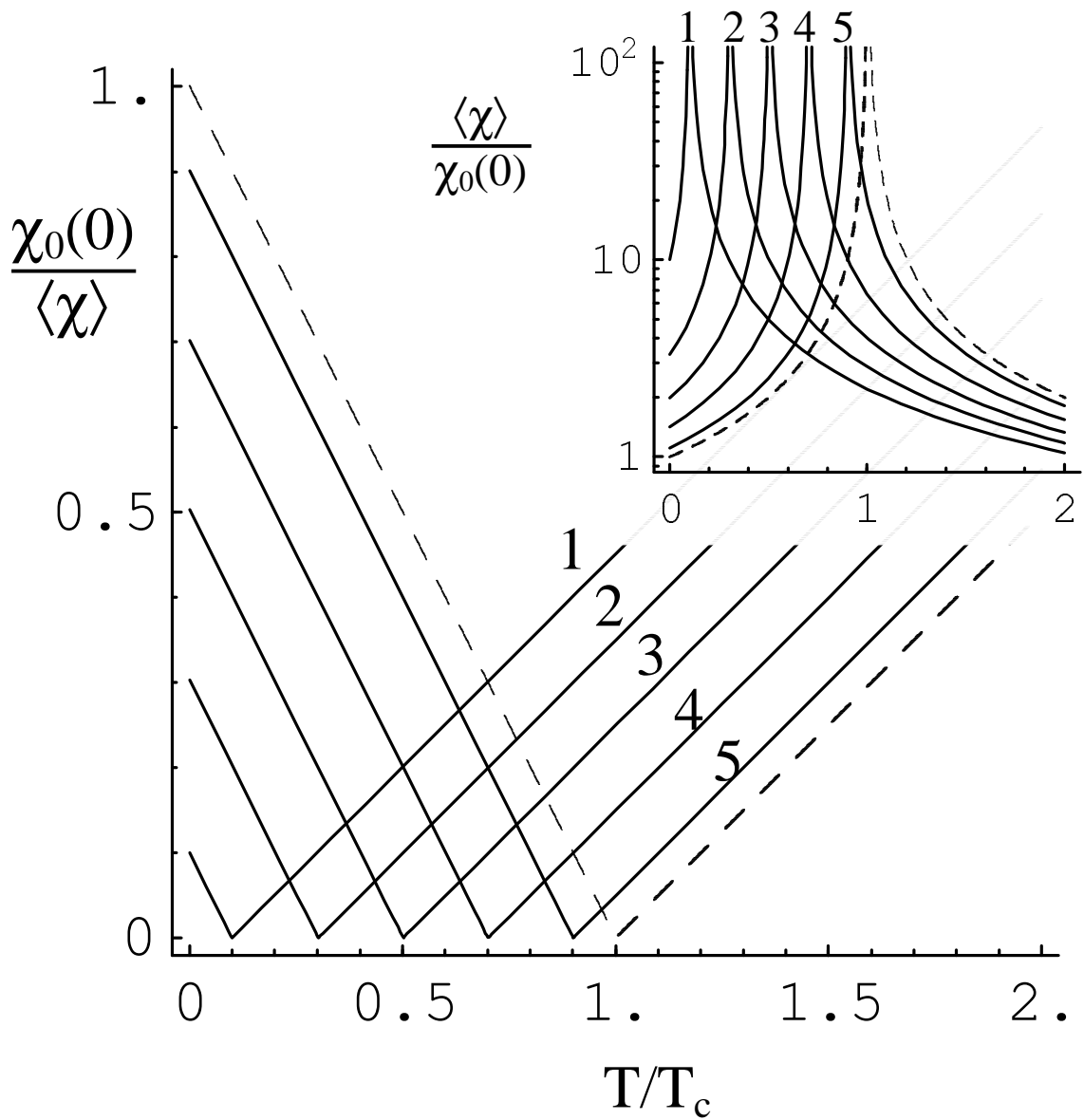
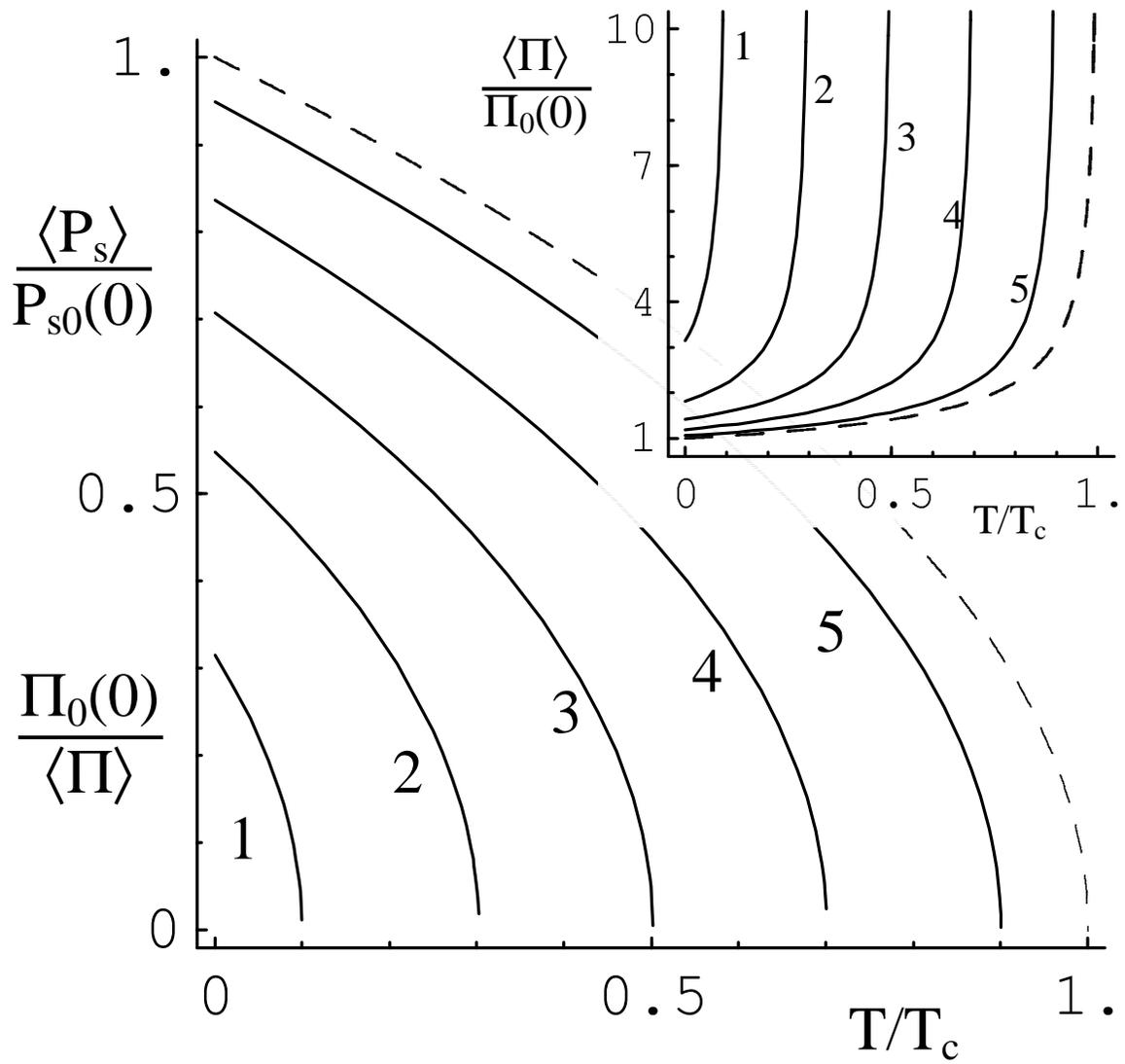
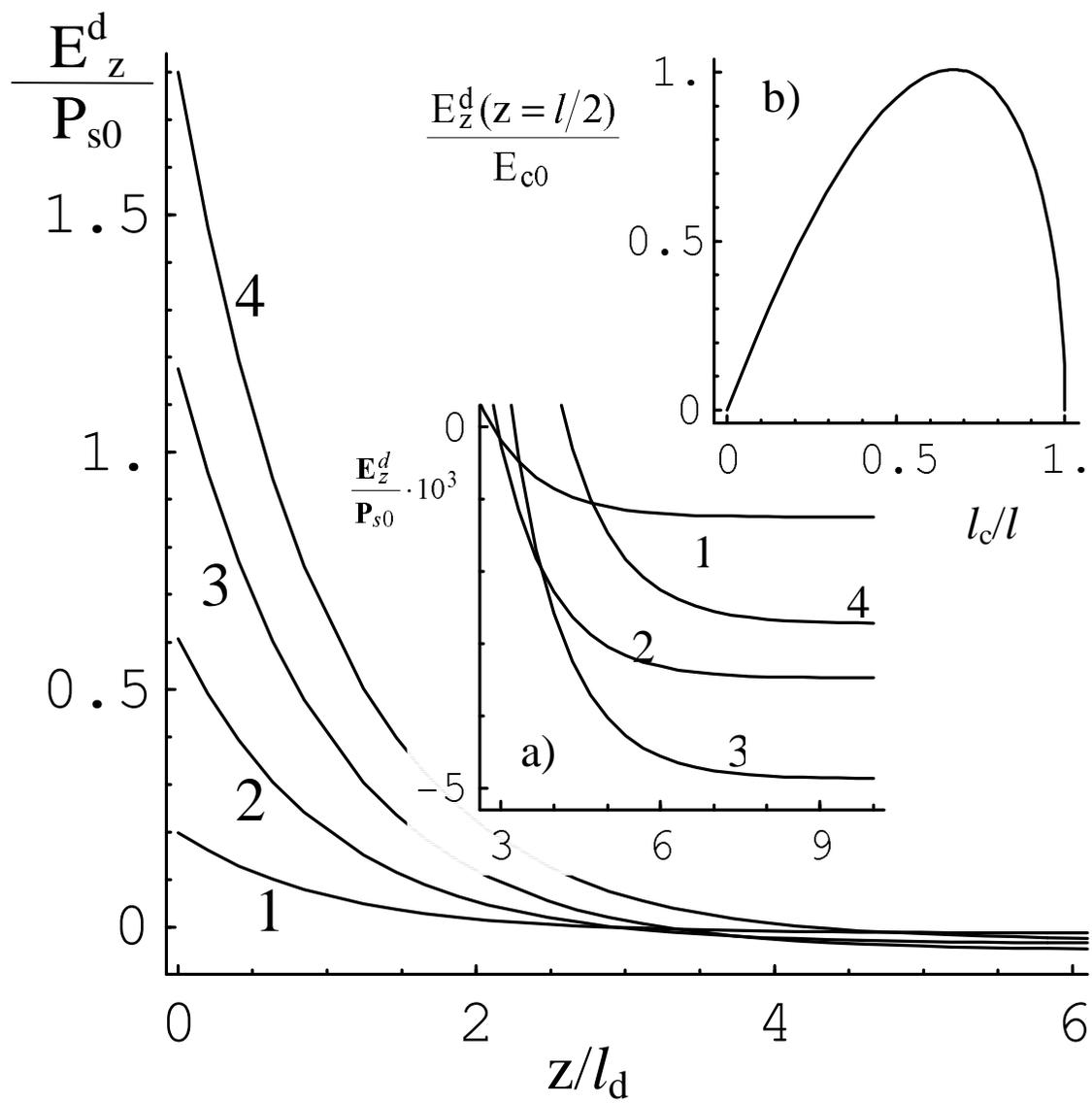
\end{document}